# Robust and Efficient Power Flow Convergence with G-min Stepping Homotopy Method

Marko Jereminov, Athanasios Terzakis, Martin Wagner, Amritanshu Pandey, Larry Pileggi
Dept. of Electrical and Computer Engineering
Carnegie Mellon University
Pittsburgh, PA

*Abstract* — Recent advances have shown that the circuit simulation algorithms that allow for solving highly nonlinear circuits of over one billion variables can be applicable to power system simulation and optimization problems through the use of an equivalent circuit formulation. It was demonstrated that large-scale (80k+ buses) power flow simulations can be robustly solved, independent of the initial starting point. In this paper, we extend the electronic circuit-based G-min stepping homotopy method to power flow simulations. Preliminary results indicate that the proposed algorithm results in significantly better simulation runtime performance when compared to existing homotopy methods.

I. INTRODUCTION

The ongoing transformation of electrical power networks towards a "smarter" grid encompasses numerous challenges. For instance, the integration of renewable energy resources within the power grid can possibly affect its performance as well as disrupt its stability at the transmission level by reversing the traditionally assumed directions of power flows [1]. Moreover, as a backbone of reliable and resilient power grid operation and planning, a robust power flow simulator that can be further scalable to joint transmission and distribution simulations [2] is a necessity in order to cope with these challenges. Hence, the revived research interest in improving the existing power flow modeling and simulation algorithms.

The industry standard for the simulation of power flow problems is based on the 'PQV' formulation [3], wherein the bus voltage magnitude and angle state variables define the steady-state response of the system, while the network constraints are modeled in terms of real and reactive power mismatch equations. The operating point of the power grid is then obtained through iteratively solving the nonlinear set of equations by numerical algorithms, such as the Newton Raphson (NR) or the Gauss-Seidel method. However, the nonlinear power balance equations are known to diverge for ill-conditioned cases operating at the edge of voltage collapse as well as suffer from lack of robustness in large-scale simulations (>50k buses), where the knowledge of a good initial guess to start the numerical algorithm can be challenging to determine. Thus, robust and scalable convergence properties of a power flow problem defined in terms of the 'PQV' formulation, as well as convergence to physical high voltage solutions remain a challenge [2].

There have been many attempts to tackle the problem of robust power flow convergence. In [4], authors propose a polynomial homotopy method to explore the power flow solution space and further locate all of the power flow solutions. This method was, however, shown to efficiently work only for small power flow test cases, and its generalization to include the real-life constraints was not explored. A different approach was introduced in [5] where a homotopy that traces a path from a DC to AC power flow solution is proposed. Soon after, [6] demonstrated that neglecting the reactive power within the DC power flow analysis represents a challenge that further causes this homotopy to trace the path of nonphysical low voltage solutions. Lastly, the Continuation Power Flow (CPF) [7] is proposed for the power flow test cases operating at the edge of voltage collapse and is applied by sequentially varying the system loading factors while resolving the power flow problem. However, the success of CPF is fully depended on the power flow solution obtained at the nominal loading, which can be as challenging to determine for the large-scale real-life test cases.

Extensive and well supported research work within the circuit simulation community has developed powerful simulation algorithms that are, today, capable of solving highly nonlinear problems with over one billion of variables. It was shown that contrary to the generalized NR damping methods that limit the NR-step vector by a constant factor [8], more robust and stable convergence properties are achieved by separately limiting each of the circuit variables within the NR-step vector [9]-[11]. More importantly, the equivalent circuit formalism [11] further allowed for development of scalable homotopy methods to ensure efficient convergence properties of large-scale circuits. For instance, one of the most popular homotopy methods, G-min stepping [12]-[14], shorts the complete circuit by connecting a large conductance from each node to the ground. The initially shorted operating point is then trivial, and the solution of the original circuit is obtained by gradually relaxing these added conductances.

Recent advances [15]-[21] in power system simulation and optimization problems have demonstrated that modeling the steady-state response of a power system in terms of current, voltage and admittance state variables allows for more robust and scalable convergence properties, governed by the application of circuit simulation techniques. For instance, the recently introduced Tx-stepping homotopy method [17] was shown to eliminate low voltage

This work was supported in part by the Defense Advanced Research Projects Agency (DARPA) under award no. FA8750-17-1-0059 for the RADICS program, and the National Science Foundation (NSF) under contract no. ECCS-1800812.

convergence issues that represents one of the drawbacks of traditional 'PQV' based homotopy methods [6]. This is achieved by "virtually" shorting every bus in the power system to a slack bus, which further enforces the homotopy to trace the path of a high voltage solution [17]. Furthermore, Tx-stepping demonstrated good scalability properties, thereby allowing the robust simulations of large-scale power flow problems (80k+ buses), as well as combined transmission and distribution analysis without loss of generality [15].

In this paper, we extend the circuit simulation G-min stepping homotopy method to power flow analysis. It is demonstrated that in contrast to the Tx-stepping method that virtually shorts the whole network to a slack bus, the G-min stepping homotopy is locally applied at each bus, and thus preserves the network topology information. Our results suggest that the initial homotopy solution is now closer to the actual steady-state grid response, which significantly improves convergence properties in terms of iteration count, when compared to the Tx-stepping and other traditionally developed 'PQV' based homotopy methods, such as CPF. Most importantly, it is shown that the proposed algorithm naturally incorporates challenging industry required power flow models, such as remote voltage control devices that require special handling within the Tx-stepping method [17]. Lastly, promising preliminary simulation result comparisons that indicate the efficiency and robustness of the proposed algorithm are presented and discussed.

## II. BACKGROUND

*A. Equivalent Circuit Formulation*

As any other electrical circuit, an electrical power system is governed by physical conservation laws, namely Kirchhoff's Current and Voltage Laws (KCL and KVL) that define the relationship between the currents and voltages within the power system. Hence, modeling and analyzing the power grid response in terms of current and voltage variables represents the most natural way of characterizing the power system in analysis problems, and as such is used in the early days of power system simulations [22] as well as first implemented on a digital computer [23]. However, these first formulations based on current and voltage state variables were later shown to represent "derivative-free" numerical algorithms [23], such as fixed-point iteration that are characterized by slow convergence properties, especially to a solution with a desired tolerance.

To address the problem of slow convergence of existing power system steady-state simulators, Tinney and Hart [3] implemented the first sparse power flow simulator based on the power mismatch formulation solved with the NR method. The power mismatch formulation was shown to reduce the size of the problem, which was valuable for the memory-constrained computers in the mid-20$^{th}$ century, and also required fewer nonlinear iterations to converge. Most importantly, it was demonstrated that the Jacobian matrix of a power mismatch formulation defined by the basic power flow models is positive definite [24], which is in contrast to the current/voltage formulation that can introduce negative eigenvalues to the Jacobian as part of the voltage-controlled PV bus model, and hence results in possible convergence issues [16]. Therefore, the power mismatch formulation became accepted as an industry standard particularly for transmission level power grids. However, positive definiteness of a Jacobian matrix does not guarantee convergence, and after the computing power of early computers increased, it was realized that as in the case with any other generic NR based algorithm, the convergence of power mismatch-based power flow is dependent on the initial starting point, particularly with the increase in size of the test cases [24].

The recently introduced equivalent circuit formulation based on current and voltage state variables has shown to overcome the challenges introduced in modeling the voltage-controlled PV nodes, by utilizing the equivalent circuit formalism and adapting the NR step limiting algorithms and homotopy methods developed within the circuit simulation community [9]-[14]. Herein, we provide a brief overview of the current/voltage formulation and the representation of the power flow problem in terms of equivalent split-circuit models.

Consider a power system given by the set of buses $\mathcal{N}$, with a set of generators $\mathcal{G}$ and load demands $\mathcal{D}$ that are subsets of $\mathcal{N}$, connected by a set of network elements, $\mathcal{T}_X$, modeled at a fundamental frequency. Furthermore, let $\tilde{V}_i = V_{R,i} + jV_{I,i}$ represent the phasor voltage of bus $i$ with its real and imaginary components respectively, while the phasor $\tilde{I}_i = I_{R,i} + jI_{I,i}$ is the current injected from the bus to the RLC transmission network comprised by the elements $\mathcal{T}_X$. It is important to note that the RLC transmission network remains linear within current/voltage formulations, while nonlinearities are introduced locally at each bus and correspond to the current injection models that constrain real and reactive powers, as well as bus voltage magnitudes [15].

Next, the complete set of governing KCL equations is split into its real and imaginary components. This is necessary to prevent the analyticity issues introduced by the complex conjugate operator in constraining the real and reactive powers (defined for $i^{th}$ bus as $P_i$ and $Q_i$) of nonlinear current injection models that would prevent the applicability of the NR algorithm. For instance, the current injection model of the $i^{th}$ bus is given as:

$$\tilde{I}_i = \frac{P_i - jQ_i}{\tilde{V}^*} \tag{1}$$

The complex current function from (1) is then split to obtain:

$$I_{R,i} = \frac{P_i V_{R,i} + Q_i V_{I,i}}{V_{R,i}^2 + V_{I,i}^2} \tag{2}$$

$$I_{I,i} = \frac{P_i V_{I,i} - Q_i V_{R,i}}{V_{R,i}^2 + V_{I,i}^2} \tag{3}$$

Lastly, in the case of a PV bus that aims to control the voltage magnitude of a bus $c$, the reactive power is treated as an additional control variable for which the voltage magnitude constraint is added as:

$$V_{R,c}^2 + V_{I,c}^2 - V_{set}^2 = 0 \tag{4}$$

where $V_{set}$ represents the voltage set point at bus $c$.

As it can be seen from (2)-(4), the governing PV bus equations are now represented by real valued functions. Hence, the first order Taylor expansion can be utilized to linearize the corresponding nonlinearities. For instance, the linearized real current from (2) is given as:

$$I_{R,i}^{k+1} = I_{R,i}^k + \frac{\partial I_{R,i}^k}{\partial V_{R,i}} \Delta V_{R,i} + \frac{\partial I_{R,i}^k}{\partial V_{I,i}} \Delta V_{I,i} + \frac{\partial I_{R,i}^k}{\partial Q_i} \Delta Q_i \tag{5}$$

To map this expression to the equivalent split-circuit representation, the term from (5), where the real current is proportional to the real voltage across it, represents a conductance by Ohm's Law, while the terms proportional to the voltage of the other circuit or the reactive power are represented by controlled current sources. Lastly, the constant terms known from the previous iteration are lumped together and represented by a constant current source, hence defining the split-circuit model as shown in Fig. 1.

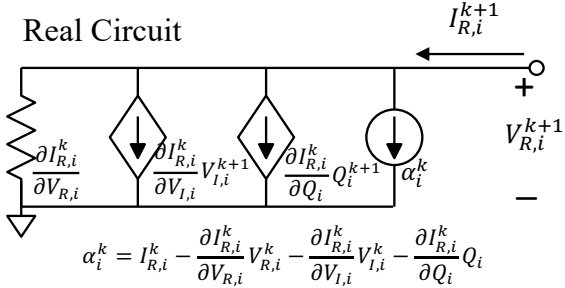

Fig. 1. Real sub-circuit of a PV bus model.

After applying the same methodology to each of the power flow models defined within the current/voltage formulation, the complete set of governing circuit equations is now linearized and can be iteratively solved until convergence to an operating point, which corresponds to NR converging to a solution. Furthermore, the equivalent circuit perspective now allows for understanding the physical meaning behind each of the components of a power flow Jacobian matrix that can be used to develop the NR-step limiting techniques based on the physical characteristics of the problem, as shown in [17]. Lastly, a detailed description of the power flow circuit models that can also include any physics-based model, such as induction motors or power electronics, as well as facilitate a joint transmission and distribution power flow analyses can be found in [15]-[21].

*B. Homotopy Methods*

Exploring and applying path tracing homotopy methods for solving general nonlinear problems has been attracting the research interest from the early days of numerical analysis. In general, a homotopy method can be mathematically defined as:

$$H(x,\mu) = (1-\mu)F(x) + \mu G(x), \quad \mu \in [0,1] \quad (6)$$

where $F(x)$ is an original nonlinear problem defined in terms of a vector of state variables $x$, while $G(x)$ represent the initial homotopy problem for which the solution can be trivially obtained. Lastly, $\mu$ is a homotopy factor that is varied in discrete steps from 1 to 0 while the homotopy problem $H(x,\mu)$ is iteratively resolved by using the solution from the previous homotopy step as an initial guess.

To further address the application of homotopy methods within an equivalent circuit formulation of a power flow problem, we provide a brief discussion on the recently introduced Tx-stepping method. Namely, Tx-stepping homotopy obtains a solution of a power flow problem [17] by embedding the homotopy factor $\mu$ to linear series and shunt network elements and transformer model, as given by:

$$G_{km} + jB_{km} = (\mu\Upsilon + 1)(G_{km} + jB_{km}) \quad (7)$$
$$t(\mu) = t + (1-t)\mu \quad (8)$$
$$\theta_{ph}(\mu) = (1-\mu)\theta_{ph} \quad (9)$$

where $G_{km} + jB_{km}$ represents the series branch admittance, $\Upsilon$ represents an admittance scaling factor, while $t$ and $\theta_{ph}$ are the transformer tap and the phase shifting angle respectively.

The power flow system equations are then sequentially solved while gradually decreasing the homotopy factor until a solution to the original power flow problem is found. Initially, the homotopy factor is set to one, and the power flow circuit is virtually "shorted" to a slack bus. Hence, the operating point of the power flow is governed by a slack bus magnitude and voltage angle and can be robustly obtained. By then gradually decreasing the embedded homotopy factor $\mu$ to zero, Tx-stepping sequentially relaxes the initially shorted power flow circuit toward its original state, while using the solution from the previous sub-problem to initialize the next homotopy decrement.

Importantly, the Tx-stepping method significantly improved simulation robustness in terms of independence on the starting point used to initialize the power flow problem. However, results suggest that its initial solution can be sometimes far away from the actual operating point, thus requiring a lot of homotopy iterations needed to converge to the solution of the original problem. Therefore, to further address the problem of simulation efficiency, we extend the idea of the circuit simulation G-min stepping homotopy method to solving the power flow problem with aim that the locally applied homotopy would provide a better starting point, and thus ensuring the shorter homotopy path to the AC power flow solution.

III. EXTENDING THE G-MIN STEPPING TO POWER FLOW SIMULATION

Obtaining the DC operating point of highly-nonlinear circuits represents the essential as well as the most challenging problem for SPICE simulations [10],[13] that can require the application of homotopy methods to ensure convergence. As one of the most popular methods used in SPICE, the G-min stepping [12],[14] connects a large conductance from every circuit node to the ground. The solution to such modified circuit is then trivial, namely zero, and the operating point of the original circuit is obtained by sequentially relaxing the connected homotopy conductances.

In the attempt to directly apply the G-min stepping to the simulation of power flow, one can realize that its direct application is not feasible mainly due to the voltage-controlled buses in the power grid. For instance, adding a large conductance from every bus to ground forces the bus voltages to zero, and thus violates the voltage-control constraints (4). Hence, in order to develop a homotopy method that does not modify the linear network models as well as acts locally at each bus, we first seek to define an initial problem that can be trivially solved to initialize the homotopy, while accurately approximating the nonlinear voltage-control PV and constant power PQ buses.

Herein, we propose a two-stage algorithm that first relaxes the nonlinear bus models in order to determine the values of homotopy admittance $G_H + jB_H$ that now by Circuit Substitution theorem [11] correspond to the relaxation error between the powers injected by the bus models of the original and relaxed power flow problems. In the second stage, the determined "error" admittances as well as the operating point of the relaxed power flow problem are used to initialize a homotopy problem $H(x,\mu)$ that is further sequentially solved until the $G_H + jB_H$ admittances are reduced to zero. The rest of the section provides a detailed description of the proposed two-stage algorithm, whose flowchart can be seen in Fig. 4.

### A. Stage I: Relaxing the power flow and determining the homotopy admittance

Consider a power system defined in terms of nonlinear current injection equations as described in II.-A. Herein, we introduce the relaxed current injection PV and PQ bus models, as well as discuss the corresponding relaxation errors that can be represented in terms of admittances and further used to initialize the homotopy in Stage II.

1. Relaxing a PV bus

To remove the nonlinearities introduced by the PV voltage-control bus, we replace it by the two linear independent current sources in real and imaginary circuits (KCL equations). Furthermore, the addition of two current state variables requires the addition of two extra equations in order to solve the set of KCL equations. Hence, we add the two linear equations constraining the voltage of a bus controlled by the PV node model (4), to the voltage set point of the original power flow problem $V_{set}$, and the angle $\theta_{DC}$ determined from the solution of a DC power flow problem, as shown in Fig. 2.

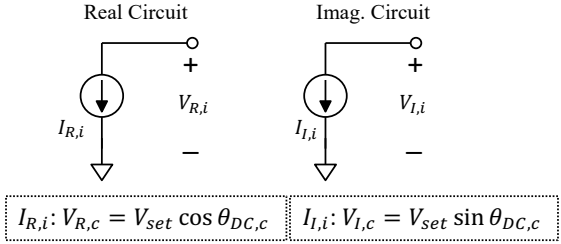

Fig. 2. Relaxed voltage control PV bus model.

Note from Fig. 2, the relaxed PV bus model naturally handles both self-control ($i = c$) and remote-control bus ($i \neq c$) models, whereas the independent current sources are connected to the *control bus*, while the voltage constraints are given for the *controlled bus*.

Importantly, DC Power Flow (DC PF) represents a simplified linear PF formulation that is solved to determine voltage angle approximations, while the grid voltage magnitudes are assumed to be equal to 1, and the system reactive powers are omitted. Hence, the real power supplied by the relaxed PV bus model from Fig. 2 will differ from set PV bus power of the original power flow problem ($P_{set}$), unless the voltage angles of AC and DC PF problems are identical. Furthermore, this "relaxation error" can be expressed in terms of admittance given at the bus voltage magnitude as:

$$G_{h,PV} = \frac{I_{R,i}V_{R,i} + I_{I,i}V_{I,i}}{V_{R,i}^2 + V_{R,i}^2} - \frac{P_{set}}{V_{R,i}^2 + V_{R,i}^2} \quad (10)$$

Furthermore, the Circuit Substitution theorem [11] states that if a voltage operating point of a node (bus) is known, all the shunt connected elements of the bus can be replaced by the independent voltage source. Furthermore, if the circuit remains unchanged, the power absorbed by the voltage source will correspond to the power absorbed by the shunt node elements. Hence, it can be implied that if an additional $G_{h,PV}$ conductance from (10) is added in parallel with the nonlinear PV bus model (see Fig. 3), the power flow solution will correspond to the solution of the relaxed problem. Therefore, this "relaxation error" from (10) further defines a homotopy conductance corresponding to a PV bus that will be sequentially stepped down in Stage II of the algorithm, while by Circuit Substitution theorem, the reactive power supplied by the relaxed model, also corresponds to the solution of the first homotopy problem, i.e. $\mu = 1$.

2. Relaxing a PQ bus

Next, we consider a remaining nonlinear PQ bus model. It should be noted that a constant PQ element can be seen as an admittance that absorbs the set real and reactive powers at a given power flow operating point. Therefore, we relax the PQ bus by replacing it with an admittance defined in terms of set powers at the nominal bus voltage magnitude ($V_B$), unless the bus voltage is controlled, in which case we evaluate the admittance at a given voltage set point.

$$G_{PQ} + jB_{PQ} = \frac{P_L - jQ_L}{V_B^2} \quad (11)$$

As we sought, the relaxed power flow problem is now represented by a linear RLC circuit and its solution can be trivially obtained. However, the real and reactive powers absorbed by the admittance from (11) will in general differ from the set real and reactive powers ($P_l$ and $Q_l$). This respective relaxation error can be defined in terms of additional conductance and susceptance values as:

$$G_{h,PQ} = \frac{P_l}{V_l^2} - G_{PQ} \quad (12)$$

$$B_{h,PQ} = -\frac{Q_l}{V_l^2} - B_{PQ} \quad (13)$$

where $V_l$ represents the relaxed power flow bus voltage magnitude operating point of bus $l$.

Finally, the Circuit Substitution theorem [11] implies that connecting the admittance from (12)-(13) in parallel with a nonlinear PQ element (see Fig. 3) ensures that the operating point of a nonlinear power flow corresponds to the solution of the relaxed problem. Therefore the "relaxation error" computed in (12)-(13) also represents the homotopy admittance related to the nonlinear PQ elements.

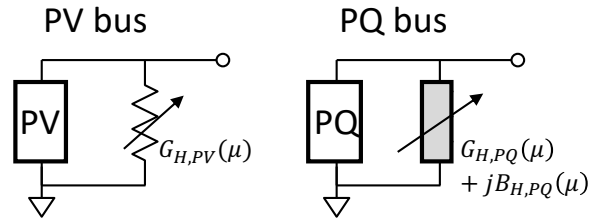

Fig. 3. Adding the homotopy conductance to nonlinear PV and PQ bus models.

### B. Stage II: G-min stepping

With the homotopy conductance and susceptance values determined from the first stage of proposed algorithm, we use the operating point of the relaxed power flow together with the $G_{h,PV}$, $G_{h,PQ}$ and $B_{h,PQ}$ to define the initial homotopy problem. Namely, by connecting the homotopy admittance in parallel to the nonlinear bus model from (2)-(3) as shown in Fig. 3, and using the real and imaginary voltage set points as well as reactive power supplied by the relaxed control bus model to initialize the nonlinear power flow. Notably, by Circuit substitution theorem, the solution to such power flow problem is equal to the already determined relaxed power flow solution. Hence, with the homotopy factor $\mu$ embedded within the computed admittances from (10), (12)-(13) as:

$$G_{H,PV}(\mu) = \mu G_{h,PV} \qquad (14)$$
$$G_{H,PQ}(\mu) + jB_{H,PQ}(\mu) = \mu(G_{h,PQ} + jB_{h,PQ}) \qquad (15)$$

the solution to the original power flow problem is determined by sequentially decreasing the homotopy admittances, while resolving the power flow problem and using the solution of the previous homotopy point to initialize the power flow at a next homotopy step.

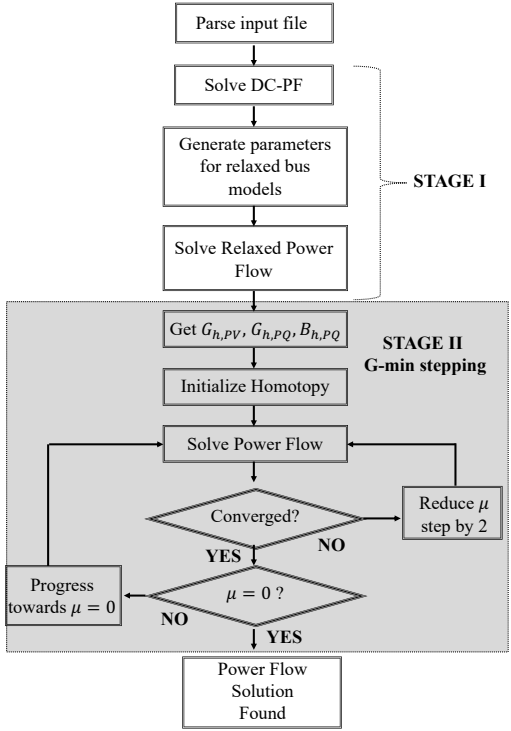

Fig. 4. Proposed 2-Stage G-min stepping algorithm.

The flowchart in Fig. 4 shows that the proposed G-min stepping from stage 2 of the algorithm applies dynamical stepping towards the original power flow problem. Namely, after the relaxed problem is solved to obtain the initial operating point as well as the respective homotopy admittances, the algorithm directly tries to obtain the solution of the original problem ($\mu \to 0$). Furthermore, if the case diverges after assigned maximum iteration count, the step of homotopy factor $\mu$ is cut back by half, until a power flow convergence is achieved. Lastly, once the homotopy factor for which the power flow converges is found, it is dynamically stepped towards zero again, until the solution of the original problem is determined.

## IV. SIMULATION RESULTS

To demonstrate the efficiency and robustness of the proposed homotopy method, the G-min stepping algorithm is implemented within a MATLAB prototype implementation of our circuit simulator SUGAR (Simulation with Unified Grid Analyses and Renewables), while the MATPOWER solver [25] was used to determine the DC power flow angle solutions. All the simulations were run on a MacBook Pro 2.9 GHz Intel Core i7, for the MATPOWER test cases including, European PEGASE test cases as well as the recently developed Synthetic cases ranging up to 70k buses [26]-[27].

First, in comparing the proposed G-min stepping and the Tx-stepping homotopy methods, it is important to note that the latter one is based on iteratively solving a relaxed problem used to start the homotopy. Namely, the Stage I of the propose algorithm solves a *linear relaxed power flow problem* that can further trivially indicate and locate the possible low voltage regions of the test case without the need of an iterative algorithm. On the other side, the Tx-stepping seeks to find an initial system response due to the large admittances used to "virtually" short the system, which can in general require a few iterations depending on the value of ϒ scaling factor.

Furthermore, in contrast to the Tx-stepping, the initial operating point of the G-min stepping homotopy is obtained without modifying the linear network elements. As our results presented in Table 1 suggest, this as a consequence has a "closer" operating point from which the G-min homotopy algorithm is started.

TABLE I. MAXIMUM ABSOLUTE VOLTAGE OPERATING POINT DEVIATION BETWEEN INITIAL HOMOTOPY AND ORIGINAL POWER FLOW PROBLEMS.

| Test Case | G-min stepping $\Delta V_m$ [p.u.] | Tx-stepping $\Delta V_m$ [p.u.] |
|---|---|---|
| Case9241PEGASE | 0.036936 | 0.279241 |
| ACTIVSg10k | 0.015694 | 0.074387 |
| Case13659PEGASE | 0.094310 | 0.259931 |
| ACTIVSg25k | 0.015153 | 0.073212 |
| ACTIVSg70k | 0.019341 | 0.095055 |

Most importantly, as we can further imply from the total iteration counts and simulation runtime comparisons presented in Table 2 and Fig. 5 respectively, the less conservative initial homotopy problem of the G-min stepping algorithm consequently provides significant improvements in terms of the simulation efficiency.

TABLE II. TOTAL ITERATION COUNT COMPARISON.

| Test Case | G-min stepping | Tx-stepping |
|---|---|---|
| Case9241PEGASE | 13 | 49 |
| ACTIVSg10k | 7 | 32 |
| Case13659PEGASE | 22 | 411 |
| ACTIVSg25k | 8 | 46 |
| ACTIVSg70k | 45 | 233 |

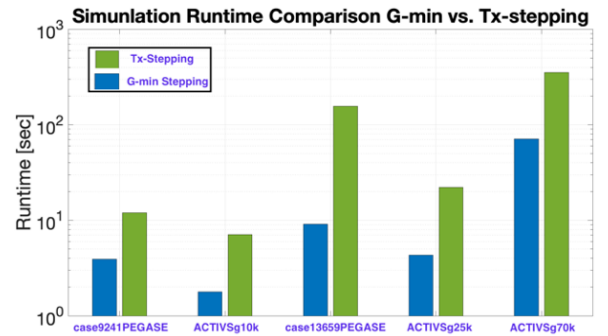

Fig. 5. Simulation runtime comparison between proposed G-min stepping and Tx-stepping homotopy methods.

Continuation Power Flow (CPF) [7] homotopy defined in terms of traditional 'PQV' formulation and power mismatch equations represents one of the more frequently used homotopy methods particularly in planning studies of power grid, where the generation and demand is increased/decreased in order to simulate extreme power grid operation conditions. Namely, to ensure the power flow convergence of the cases with scaled loading factors, the loading factor is sequentially varied starting from the power flow solution obtained for the initial loading of the test case, while resolving the power flow until the desired target loading is reached.

Therefore, as a second experiment, we compare the proposed G-min stepping with the MATPOWER [25] implementation of the CPF algorithm. For this study we consider four synthetic test cases [26] representing:

1. South Carolina (500 buses)
2. Texas – ERCOT (2000 buses)
3. Western Interconnect – WEC (10,000 buses)
4. North Eastern region of USA (25,000 buses)

and further increase the respective loading factors by **25%**. The CPF is run for the *default parameters* in MATPOWER, and the obtained runtimes for each of the four examined test cases are compared in Fig. 6 with the runtimes obtained from solving the power flow with G-min stepping algorithm.

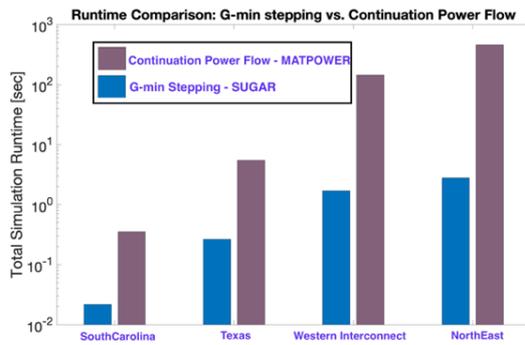

Fig. 6. Simulation runtime comparison between proposed G-min stepping and Continuation Power Flow method. Note the significant improvement (order of magnitude) in simulation runtime efficiency.

## V. CONCLUSONS

In this paper, we extended the circuit simulation G-min stepping homotopy method to solve the power flow problem. The proposed homotopy method is demonstrated to naturally incorporates the challenging, industry required power flow models, such as remote voltage control devices. Furthermore, with the presented preliminary results that indicate significant improvements in the simulation efficiency of the proposed G-min stepping in comparison to existing homotopy methods, it is important to emphasize the increase of total iteration count once a homotopy method is applied. Importantly, as shown in [15], the convergence of the well-conditioned power flow cases can be robustly obtained using the circuit simulation NR-step limiting techniques and is generally not dependent on the application of a homotopy method. However, as in the case of circuit simulators based on SPICE, efficient and robust homotopy methods *should represent an important component of every power flow simulator*. The component that can be called for in the worst-case scenarios, when the power flow solution cannot be obtained with any of the NR step limiting techniques or during the convergence to a physically meaningless solution, i.e. low voltage solution.